\documentclass[twocolumn,prd,floatfix]{revtex4}

\usepackage{color}
\usepackage{tabularx}
\usepackage{epsfig}
\usepackage{amsmath}
\usepackage{amssymb}
\usepackage{bm}
\usepackage{graphicx}
\usepackage{multirow}
\usepackage{physics}
\usepackage{epsfig}

\begin{document}

\title{ Complementarity-Entanglement Tradeoff in Quantum Gravity}

\author{Yusef Maleki$^1$ and  Alireza Maleki$^{2}$}
\affiliation {$^1$Department of Physics and Astronomy, Texas A\&M University, 
College Station, Texas 77843-4242, USA\\
$^{2}$ Department of Physics, Sharif University of Technology, Tehran, Iran}
\date{\today}

\begin{abstract}

Quantization of the gravity	remains  one of the most important, yet extremely illusive, challenges at the heart of modern physics. Any attempt to resolve this long-standing problem seems to be doomed, as the route to any direct  empirical evidence (i.e., detecting gravitons) for shedding light on the quantum aspect of the gravity  is far beyond the current capabilities. Recently,  it has been discovered that  gravitationally-induced entanglement, tailored in the interferometric frameworks, can be used to witness the quantum nature of the gravity. Even though these schemes offer promising tools for investigating quantum gravity, many fundamental and  empirical aspects of the schemes are yet to be discovered. Considering the fact that, beside quantum entanglement, quantum uncertainty and complementarity principles are the two other foundational aspects of quantum physics, the quantum nature of the gravity needs to manifest all of these features. Here,
we lay out an interferometric platform for  testing these three nonclassical aspects of quantum mechanics in quantum gravity setting, which
connects gravity and quantum physics in a broader and deeper context. As we show in this work, all of these three fundamental  features of quantum gravity can be framed and fully analyzed in an interferometric scheme.

\end{abstract}

\maketitle

\section{Introduction}
Gravity,  as one of the four building block forces of the universe,  is the most sensible of the forces that we deal with in our every day lives. However, it also is considered as the most controversial one in modern physics.
The general theory of relativity for gravitation, which was formulated by Einstein in 1915, aside from its thundering triumphs in explaining the universe, encounters important  issues at the high energy levels, such as black holes information paradox and singularity problem \cite{biswas2012towards,brustein2014origin}.  These issues arise from the fact that there is no known way to reconcile  gravity with the other  three  forces of the nature. It seems that to rectify these issues one needs to integrate quantum physics and gravity in a unified frame, which remains as one of the most important unsettled issues in physics.  

 Therefore, there has been an ongoing attempt to establish a quantum theory of gravity, leading to development of different approaches such as string theory and quantum loop gravity \cite{carlip2001quantum,kiefer2007quantum,rovelli2008loop}. On the other hand, considering the aforementioned  challenges, the search for alternative theories such as theories of emergent gravity, instead of quantum gravity has also  attracted a great attention \cite{padmanabhan2015emergent,verlinde2017emergent}.
 The main hurdle in  establishing a quantum theory for gravity is associated with  the weakness of the gravitational interaction. In other words, the quantum properties of gravity  becomes sensible only in the range smaller than the plank scale, requiring experiments with energies well beyond the scales that are within the near-future capabilities. This fact makes it impossible to discern a reliable route to the problem and to discriminate between various developed models of quantum gravity \cite{alfaro2005quantum}.  Therefore, laying out some empirically feasible methods for studying and testing the quantum nature of gravity  is of utmost importance.

In recent years, there has been an increasing interest in the quantum-information-theoretic approaches to the study of relativistic and gravitational systems \cite{peres2004quantum,lunghi2013experimental,maleki2020speed}. 
 Recently, as a novel approach for probing the quantum nature of the gravity, it has been discovered that gravitationally-induced entanglement can indeed serve as a witness of quantumness of the gravity  \cite{bose2017spin,marletto2017gravitationally}. As is demonstrated in  \cite{bose2017spin,marletto2017gravitationally}, the  detection of entanglement, attained by gravitational interactions of massive particles, can be considered as a sufficient criteria for quanta of the gravitational field. As was demonstrated, such entanglement generations could be tailored in interferometric schemes, and considering the rapid progress in quantum information and interferometry technologies, the experimental feasibility of these approaches are within the sight.

Even though entanglement is central to the Hilbert space structure of composite quantum systems, the quantum feature of the systems is in no way restricted to the entanglement. In fact,
in quantum physics, Heisenberg's uncertainty principle \cite{heisenberg1985anschaulichen},  Bohr's complementarity principle \cite{bohr1928} and quantum correlations in composite quantum systems  are the most fundamental aspects of  systems, such that the entire fabric of the quantum weirdness can be captured by these fundamental characteristics \cite{zubairy2020quantum}.
Therefore, it is imperative to search for all these three building blocks of quantum mechanics in quantum nature of the gravity, in a feasible experimental setup.

As an other fundamental aspect of quantum physics, Bohr's complementarity principle \cite{bohr1928}  provides one of the most fundamental aspects of the nature, which explains that the two mutually exclusive attributes, such as the waviness and the particleness, can both be imprinted in quantum systems, such that measuring one feature prohibits its dual feature to be exhibited \cite{bohr1928,zubairy2020quantum}. As an example, in the case of a single photon passing through an interferometer, the particleness of the photon is embedded in the path predictability  of the photon, while the waviness is encoded in the visibility of the interference pattern on screen \cite{scully1994}. The quantitative notion of such a discipline was first introduced by Wootters and Zurek  in 1979 \cite{Wootters1979}, which lead to a mathematical description in form of an 
inequality as $\mathcal{P}^2+\mathcal{V}^2\leq 1$, where $\mathcal{P}$ represents the predictability  of a quantum system, which contains the path information and indicates a quantity for particleness, and $\mathcal{V}$ stands for the visibility of interference pattern, measuring the  waviness of the system  \cite{Glauber1986,Mandel1991,Jaeger1993,Englert1996}. 

In an interesting attempt, it has been shown that the wave-particle duality has a relationship with the entanglement in the system \cite{jakob2007complementarity,deMelo,Qian2018}. This unification of the duality and entanglement in double slit (two-qubit) analyses provided an important relation as $\mathcal{P}^2+\mathcal{V}^2+\mathcal{C}^2= 1$ \cite{jakob2007complementarity,deMelo,Qian2018}, in which  $\mathcal{C}$ is a measure of entanglement known as concurrence \cite{Wootters1998}.
More recently, an interesting geometrical correspondence  between this relation and stereographic projection of $S^7$ geometry was also formulated \cite{Maleki},  
which gives a full geometrical proof for the duality-entanglement relation \cite{Maleki}.

In  this paper, we put forward an experimentally feasible platform that enables testing Heisenberg's uncertainty principle,  Bohr's complementarity principle and quantum entanglement as the nonclassical aspects of quantum mechanics in quantum gravity framework,
connecting gravity and quantum physics in a broader and deeper context. As one important aspect of our study, we show that all of these three fundamental  charactristics of quantum gravity can be framed and tested in an interferometric scheme.

 The paper is organized as follows. In Section \ref{1QG}, we introduce the quantum gravitational interaction potential and discuss its fundamental implications. In Section \ref{2statecomplementarity}, we briefly address the entanglement-complementarity relation and its main features. In Section \ref{Nstatecomplementarity}, we put forward the gravitaionally induced complementarity and entanglement analyses in an interferometric quantum superposition scenario.
 In Section \ref{Bell}, we briefly analyze
Bell inequality which provides a practical way for the detection of entanglement in quantum systems.
In Section \ref{Uncertainty}, we lay out the discuss of the uncertainty principle in the context of graviationally induced quantum phases.
In Section \ref{experimental}, we discuss the experimental feasibility of the gravitaionally induced entanglement and its analyses provided in this work. We finally provide a short summary and conclude in \ref{Conclusion}.

 \section{Main aspects of quantum gravity}
   \label{1QG}
 Here, we briefly consider some important aspects of the gravitationally induced phase, and the significance of  potential that results into such an entanglement. To this end, we start with the action of the gravitational field as \cite{hobson2006general}
 \begin{equation}
 S=\int d^4x \sqrt{-g} [R+\mathcal{L}_{m}],
 \end{equation}
 where $g$ is the determinant of metric $g_{\mu\nu}$, and $R$ is Ricci scalar which depends on the derivations of the metric, and is  related to the Ricci tensor by $R=g^{\mu\nu}R_{\mu\nu}$. Also, $\mathcal{L}_{m}$ represents the matter part of the action.
 
 Now, taking the variation over metric one can reach to the Einstein field equation \cite{hobson2006general}
 \begin{equation}
 R_{\mu\nu}-\frac{1}{2}Rg_{\mu\nu}=8\pi G T_{\mu\nu},
 \end{equation}
 where $T_{\mu\nu}$ is the stress-momentum tensor. To make the quantum nature of the gravity manifest, we study the perturbation $h_{\mu\nu}$ on the background metric of  Minkowski space-time $\eta_{\mu\nu}$. Therefore, the entire metric can be written as 
 $g_{\mu\nu}=\eta_{\mu\nu}+h_{\mu\nu}$. Considering the weak field limit, after expanding the metric and omitting the orders higher than quadratic terms and choosing the harmonic gauge we obtain \cite{zee2010quantum}
\begin{align}
\nonumber
 S_{WF}&= \frac{1}{2}\int d^4 x ((-\frac{1}{32\pi G}
 \\ 
 &\times(\partial_{\lambda}h^{\mu\nu}\partial^{\lambda}h_{\mu\nu}-\frac{1}{2}\partial_{\lambda}h\partial^{\lambda}h)+h^{\mu\nu}T_{\mu\nu}).
 \label{2rdActionGW}
 \end{align}
 Now, considering $h$ as a field, the first two terms refer to the free field of the gravity and the third term indicates the interaction of the gravity with the matter.  The free gravitatioal field can be written as \cite{zee2010quantum}
 \begin{equation}
 h^{\mu\nu}= \frac{1}{(2\pi)^3} \int\frac{ d^3 k}{\sqrt{\omega}} \sum_{\sigma=\pm2} (e^{\mu\nu}_{\sigma}(\vec{p}) a_{\sigma}(\vec{p})e^{i\vec{p}.\vec{x}}+h.c.),
 \label{hfurier}
 \end{equation}
 where $\sigma$ denotes  the spin of the graviton and   $e^{\mu\nu}$ is the polarization tensor with the properties $e_{\sigma}^{\mu\nu} e_{\sigma'}^{\mu\nu\star}=2\delta_{\sigma\sigma'}$.
 
 As a result, we can calculate the propagator and the  scattering process of two masses. To illustrate, in one loop level, after Fourier-transformation of scattering amplitude in momentum space, and in the non-relativistic limit, the potential can be obtained as \cite{radkowski1970some,kirilin2002quantum,bjerrum2003quantum}
 \begin{equation}\label{potential}
 V(r) = -\frac{Gm_1m_2}{r} (1 +3\frac{G(m_1+m_2)}{rc^2} + \frac{41 G\hbar}{10\pi r^2 c^3}).  
 \end{equation}
 The second term is the General Relativistic correction, and the third term is related to the quantum gravity correction to the classical potential. These two terms are extremely small compared to the first term. More specifically the last terms is quite negligible, compared to the first two terms; therefore, the detection of quantum nature of the gravity does not seem to be feasible if one tries to detect it by using the last term of this equation. This, in fact, shows why observation of the quantum mechanical nature of  gravity is such a difficult task.  
 
Considering the extremely small contribution of quantum part in the potential in Eq. (\ref{potential}), the substantially important quantum is whether it is possible to observe the quantum nature of the gravity through the fist term (the Newtonian approximation)  of the potential? The answer to this question is surprisingly, yes. This could be achieved by a purely information-theoretic approach to the  gravity. In particular,
 as recently was discovered, an interesting route to this goal is the gravity induced entanglement, which circumvents the challenges of the weak strength of the quantum contribution to the gravitational interactions. The subtleness of the idea lies in the fact that the induced entanglement can emerge even at the low energy limits, where we can approximate the potential with its first term, i.e., the Newtonian approximation \cite{marshman2020locality,andersen2019quantum}. Interestingly, this entanglement-assisted analyses of the quantum nature of gravity offers an advantage for the accessibility  of the induced quantumness in an empirical set up, compared to the effect that are directly proportional to Planck constant as appears in last term of the potential in Eq. (\ref{potential}). In the next sections we discuss about this remarkable approach toward the investigating of quantum gravity.

The important role of entanglement is inherited by the fact that  if the gravitational interaction can generate entanglement among two masses, the graviational field itself should be of quantum nature. In other words, we assume that two masses interact via gravity, in which the interaction generates entanglement.  In this setting,  gravity acts as a mediator of the entanglement between two masses. The interaction is induced by the exchange of  gravitons as the mediator of the entanglement. If the entanglement is generated then graviational field needs to be coupled quantum mechanically to each test mass in order for the generation of the entanglement. This is due to the fact that no classical mediator can generate entanglement, as was shown in \cite{krisnanda2017revealing}. A similar conclusion can be drawn from LOCC theorem which states that no local operation and classical communication (LOCC) can increase the entanglement \cite{nielsen2002quantum}. Therefore, starting with two disentangled masses, the entire system remains disentangled unless some quantum mechanical interactions are applied. As an important conclusion of this fact, detection of entanglement in such a setting is suffix to conclude that the gravity is of quantum nature.

 \section{ Two-qubit Complementarity-entanglement relation}
 \label{2statecomplementarity}
 
 The space of a two-qubit state is the product of Hilbert space of each qubit denoted by $H^\mathbb{C}_1 \otimes H^\mathbb{C}_2$, offering a 4-dimensional Hilbert space for the system. Considering the fact that each Hilbert space, in this setting, can be spanned by two orthogonal basis $\{\vert 0\rangle, |1\rangle\}$, the basis of two-qubit Hilbert space can be written as $\{\vert 00\rangle, |01\rangle,\vert 10\rangle,\vert 11\rangle\}$, where $\vert ij\rangle$ denotes the composite basis as $\vert i\rangle \otimes \vert j\rangle$, with $i,j=0,1$.
  Therefore, the general form of a pure two-qubit state can be expressed as
  \begin{align}
\vert \psi\rangle=\alpha_{0}\vert 00\rangle+\alpha_{1}\vert 01\rangle+\alpha_{2}\vert 10\rangle+\alpha_{3}\vert 11\rangle.
  \label{2state}
 \end{align}
Interestingly, we can encode any two-state system, such as the paths in  Young double-slit experiments  or Mach Zehnder interferometer as a correspondence to a two-qubit system. 

\begin{figure}
	\includegraphics[width=3 in]{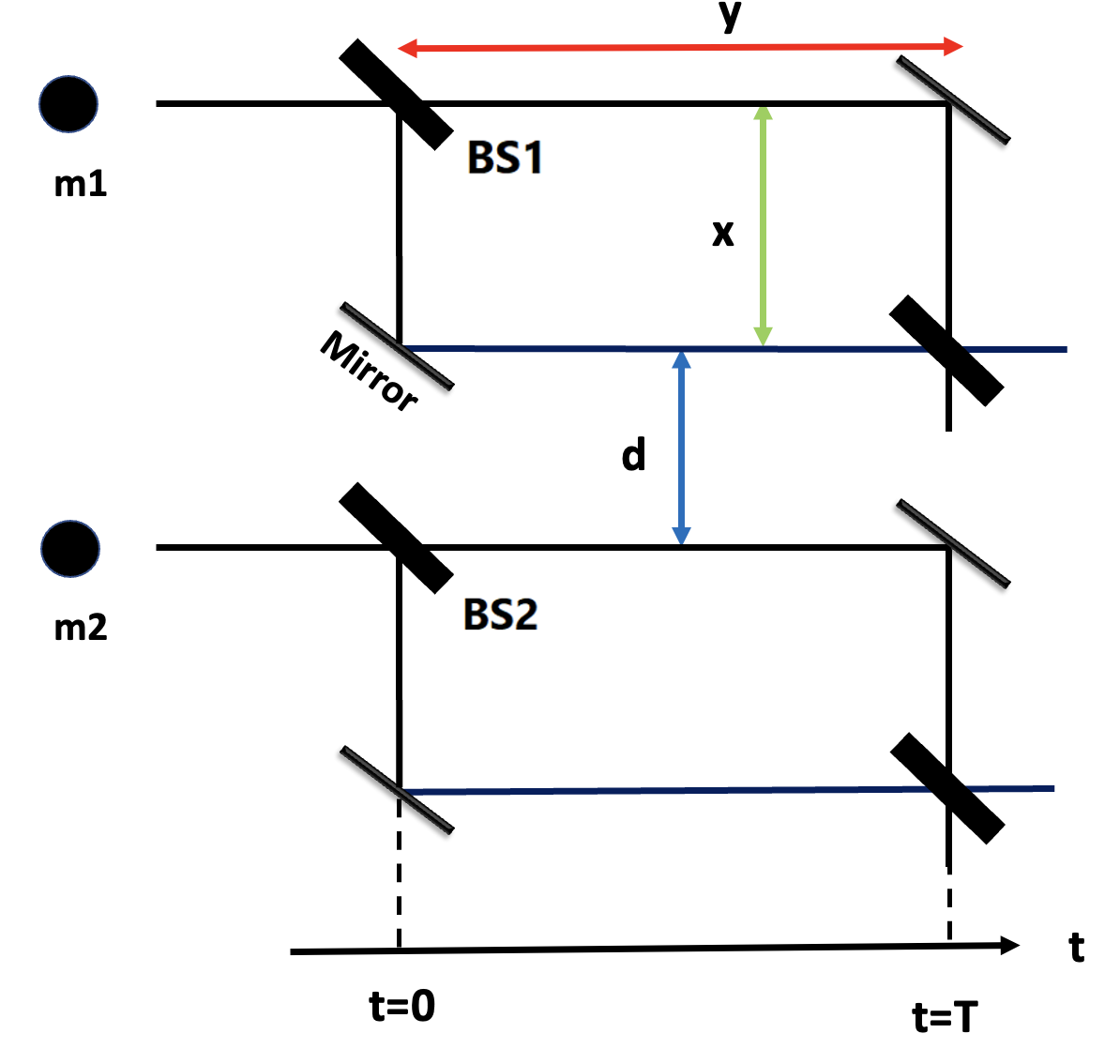}
	\centering
	\caption{Experimental set-up of interferometer for gravitationally induced entanglement. The masses $m_1$ and $m_2$
 undergo independent Mach Zehnder type interference, and interact with each other by gravity. $BS_i$ ($i=1,2$)
indicates a beam splitter which is characterized by $(r_{i},t_{i})$. The second beam splitter in each  interferometer is assumed to be a 50:50 beam splitter.
	}
	\label{fig1}
\end{figure}

To characterise the set up for quantum analyses of the gravity  in this work, we consider two massive particles, each subjected to a Mach Zehnder interferometer as depicted in Fig. \ref{fig1}. We can  assign to the upper path and the lower path states in the interferometer two orthogonal states $|u\rangle$ and $|d\rangle$, respectively.
This encoding is equivalent to $|1\rangle$ and $|0\rangle$ in Eq. (\ref{2state}). Hence, assigning the basis  $|u_i\rangle$ and $|d_i\rangle$  for the $i$th interferometer ($i=1,2$), the system of the two interferometers is described by the state of the form
 \begin{align}
 	\vert \psi\rangle=\alpha_{0}\vert u_1u_2\rangle+\alpha_{1}\vert u_1d_2\rangle+\alpha_{2}\vert d_1u_2\rangle+\alpha_{3}\vert d_1d_2\rangle
 	\label{3state}.
 \end{align}

 With this description, we can study the duality-entanglement concept in a two-qubit system as shown the setup in  Fig. 
 \ref{fig1}. In our analyses of the wave-like and particle-like features, without loss of generality, we  consider the first interferometer (first mass). Naturally, a similar discussion is also valid for the second interferometer.
 
 Now, wave-like feature in this system (first mass) is quantified by the visibility, which is determined by  \cite{Qian2018,deMelo, jakob2007complementarity,Maleki}
\begin{align}
\mathcal{V}=\frac{p_{D}^{max}-p_{D}^{min}}{p_{D}^{max}+p_{D}^{min}}, 
 \end{align}
where $p_D$ is the probability of detecting the system after the second beam-splitter of the first interferometer, and the upper indexes indicate the maximum and minimum of the probability. Applying this relation to the quantum state in Eq. (\ref{3state}), the visibility reduces to \cite{Maleki}
\begin{align}
\mathcal{V}=2 \mid {\alpha_{2}^*}\alpha_{0}+{\alpha_{3}^*}\alpha_{1}\mid.
 \label{vis2S}
 \end{align}
In a similar vein, the predictability is determined as  \cite{Qian2018,deMelo, jakob2007complementarity,Maleki}
\begin{align}
\mathcal{P}=\frac{ \mid p_{u}-p_{d} \mid}{ \mid p_{u}+p_{d} \mid}, 
 \label{disting}
 \end{align}
where the parameters $p_{u}$ and $p_{d}$ are the probabilities of finding first particle in each of the chosen paths (i.e., the probability of finding the first mass in the upper or the lower arm of its corresponding interferometer). Hence, from this equation the particleness of system can be attained as \cite{Maleki}
 \begin{align}
\mathcal{P}= |{(|\alpha_{0}|}^2+{|\alpha_{1}|}^2)-({|\alpha_{2}|}^2+{|\alpha_{3}|}^2)|.
 \label{disting2S}
 \end{align}
On the other hand, the amount of entanglement in the above system  can be obtained using concurrence, which is defined through $R$ matrix such that $R=\sqrt{\sqrt{\rho} \bar{\rho}\sqrt{\rho}}$, with $ \bar{\rho}=(\sigma_x \otimes \sigma_x) \rho^{\star}  ( \sigma_x \otimes \sigma_x)$. By organizing the eigenvalues of the  $R$ matrix  in decreasing order, the concurrence is defined as \cite{Wootters1998}
 \begin{align}
\mathcal{C}=\text{max}\{0,\lambda_0-\lambda_1-\lambda_2-\lambda_3\},
 \end{align}
In which the eigenvalues of matrix $R$ is denoted by  $\lambda_i$ in the deceasing order.
In the case of the state in Eq. (\ref{3state}), the concurrence is given by \cite{Maleki}
\begin{align}
\mathcal{C}=2 \mid \alpha_{0}\alpha_{3}-\alpha_{1}\alpha_{2}\mid.
 \label{concor2S}
 \end{align}
 Recently, it was  realized that concurrence plays a significant role in a quantum complementarity setting, such that \cite{Qian2018, deMelo}
 \begin{align}
\mathcal{P}^{2}+\mathcal{V}^{2}+\mathcal{C}^{2}=1.
 \label{complementarity2s1}
 \end{align}
 This relation shows that entanglement can indeed control the duality in quantum system, hence, the full description of the system entails wave-particle-entanglement triality relation as above, rather than the duality description alone.
 This relation also can be proven geometrically as outlined in Ref. \cite{Maleki}, where it was  shown that complementarity principle can indeed be analysed from a completely geometric perspective.


  \section{gravitaionally induced complementarity and entanglement}
  \label{Nstatecomplementarity}
As we discussed earlier, the ultimate theory of quantum gravity should enable us to express gravity as a superposition of different states \cite{marshman2020locality,andersen2019quantum}. This is a main feature of quantum gravity which makes it different from classical case and could be used in finding an experimental approach to test the theory in a table-top experiment. Following this line of thoughts,
recently, it was found that quantum nature of gravity can be tested in the light of entanglement generated by the gravity \cite{bose2017spin,marletto2017gravitationally}.
   In these interesting works, it was shown that interferometric setup enables testing the existence of the gravitationally-induced entanglement between two masses.
   
   Here, we lay out a rather general framework to consider quantum foundation of gravity. To this end,
    we assume  two massive particles, $m_1$ and $m_2$, each subjected to an interferometer as depicted in  Fig. \ref{fig1}.   We also assume that, in  the first interferometer the first beam splitter, which particle encounters, is characterized by $(r_{1},t_{1})$, where $r_i$ and $t_i$ represent the refectivity and transmissivity parameters of the $i$th beam splitter. Similarly, in the second  interferometer its first beam splitter is characterized by $(r_{2},t_{2})$. For simplicity, we could assume that the arms which is perpendicular to the initial direction of motion of the particles in interferometer are negligible in comparison to the parallel arms \cite{bose2017spin,marletto2017gravitationally}. Therefore, the initial state of the two masses after passing through the first beam splitters can be expressed as the product of states of the first and second particle paths, which is given by
  \begin{align}
 |\psi(t=0)\rangle=(t_1|u_1\rangle+r_1|d_1\rangle)\otimes(t_2|u_2\rangle+r_2|d_2\rangle).
 \label{1ifnt}
 \end{align}
Note that the setting considered here is much more general compared to the previous studies \cite{bose2017spin,marletto2017gravitationally} where only 50:50 beam splitters were taken into the consideration.

  \begin{figure}
 	\centering
 	\includegraphics[width=3.5 in]{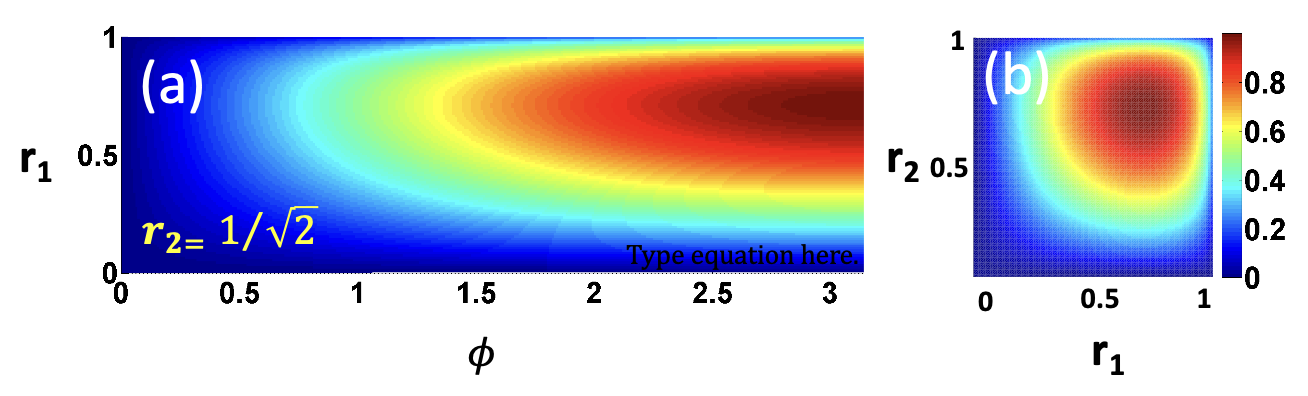}
 
 	\caption{The concurrence of the system as a function of different parameters in the setup: (a) shows the dependence of concurrence on the gravitaionally induced phase and  reflectivity of the first  beam splitter in the first interferometer. (b) depicts the concurrence as a function of reflectivity of the first beam splitters in each interferometer, where the gravitaionally induced phase is set to  $\phi=\pi$.}
 	\label{fig2c1}
 \end{figure}

  \begin{figure}
  	\centering
 	\includegraphics[width=3.6 in]{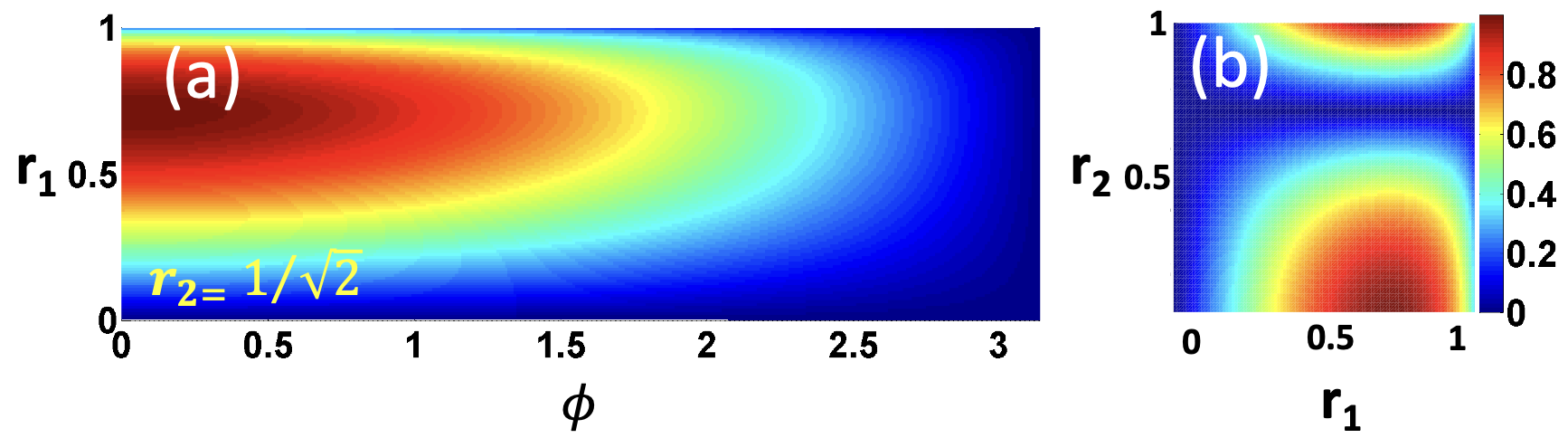}
 	\caption{The visibility of the system as a function of different parameters in the setup: (a) the dependence of visibility on the gravitaionally induced phase and  reflectivity of the first beam splitter in the first interferometer. (b)  visibility as a function of reflectivity of  first beam splitters in each interferometer, assuming that the gravitaionally induced phase is  $\phi=\pi$.
 	}
 	\label{fig2v1}
 \end{figure}

 \begin{figure}
	\includegraphics[width=3.6 in]{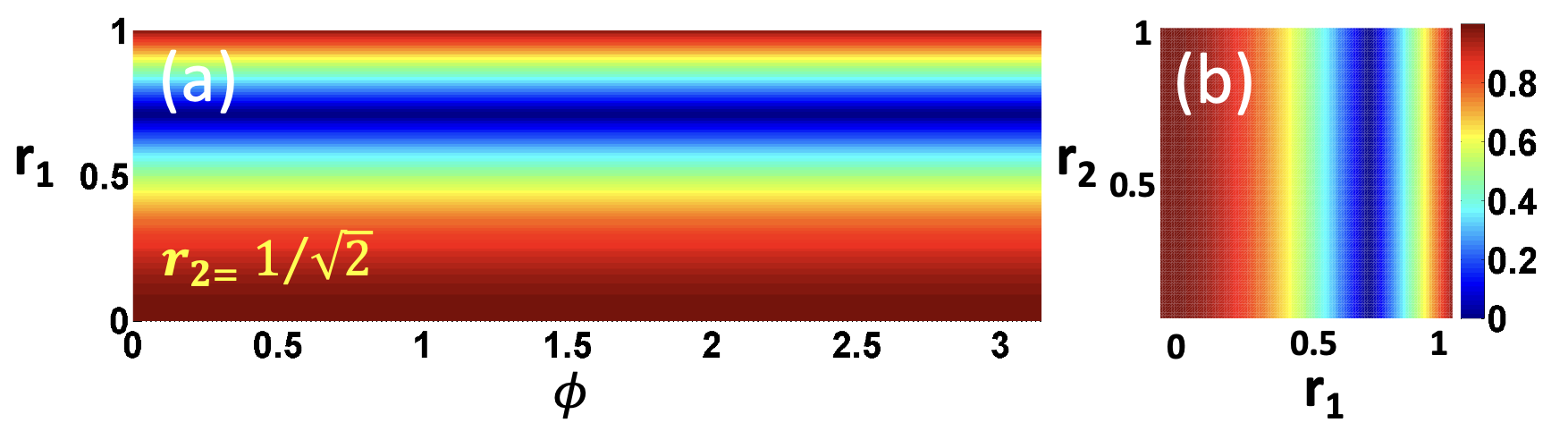}
	\centering
	\caption{The predictability of the system as a function of different parameters in the setup: (a)   dependence of predictability on the gravitaionally induced phase and  reflectivity of the first beam splitter in the first interferometer. (b) predictability as a function of reflectivity of first beam splitters in each interferometer, where the gravitaionally induced phase is fixed to  $\phi=\pi$.}
	\label{fig2p1}
\end{figure}

The interaction of a quantum particle with gravitational field results in an induced phase, which its first experimental demonstration was attained in a famous experiment by  Colella,  Overhauser and  Werner (which is known as the COW experiment) in 1975 \cite{colella1975observation}. Now, similar to this experiment, we consider the setting where the gravitational field can indeed induce phase shift on the quantum systems \cite{bose2017spin,marletto2017gravitationally}. In this setting gravity induces a phase shift and decouples from the system once the phase shift is attained \cite{bose2017spin,marletto2017gravitationally}. Therefore, the initial state of the systems given by $|\psi(t=0)\rangle$ will evolve into
\begin{align}
\nonumber
|\psi(t=T)\rangle&=r_1r_2 e^{i\phi_1}|d_1\rangle |d_2\rangle+r_1t_2 e^{i\phi_2}|d_1\rangle |u_2\rangle\\ 
&+ t_1r_2 e^{i\phi_3}|u_1\rangle |d_2\rangle 
+t_1t_2 e^{i\phi_4}|u_1\rangle |u_2\rangle,
\label{3ifnt}
\end{align}
before entering the second beam splitter in each interferometer.
Here, the phases are given by
\begin{align}
\nonumber
	\phi_1&=G\frac{m_1m_2}{\hbar d}T,  \qquad \phi_2=G\frac{m_1m_2}{\hbar (d+x)}T, 
	\\
	\phi_3&=G\frac{m_1m_2}{\hbar (d-x)}T, \qquad \phi_4=\phi_1.
\end{align}
In which $T$ is the time duration of gravitational interaction in each arm  and $x$ is the width of interferometer.

Hence, this state at hand,  we can calculate the entanglement using Eq. (\ref{concor2S}), from which we obtain
\begin{eqnarray}
\mathcal{C}=2r_1r_2t_1t_2 \mid 1- e^{i(2\phi_1-\phi_2-\phi_3)}\mid.
\label{2C}
\end{eqnarray}
This immediately results in
 \begin{align}
 \mathcal{C}^2=8(r_1r_2t_1t_2)^2 (1- \cos(\phi)),
 \label{cm1}
 \end{align}
in which $\phi=2\phi_1-\phi_2-\phi_3$. It is easy to check that when $\phi=2n\pi$ (for $n$ being an integer number) the entanglement is zero, while it becomes maximum when $\phi=n\pi$ ($n$ being an odd number).

The behavior of this equation is presented in Fig. \ref{fig2c1}. In  Fig. \ref{fig2c1}(a),  the dependence of the concurrence on the gravitaionally induced phase and  reflectivity of the first beam splitter in the first interferometer is illustrated. In Fig. \ref{fig2c1}(b) also, the  concurrence as a function of reflectivity of  first beam splitters in each interferometer is plotted, in which the gravitaionally induced phase is set to be $\phi=\pi$.  In both cases when we choose the reflectivity of beam splitters as $r_1=r_2=1/\sqrt{2}$, the entanglement becomes maximum.

We note that the entanglement generated in this setting is indeed induced by gravitational field.
 Here, gravity acts as a mediator of the entanglement, and if the entanglement could be observed we can conclude that the gravity is of quantum nature. This is due to the fact that no classical mediator can generate entanglement \cite{krisnanda2017revealing}. 
 
 Next, we consider the visibility of the interference in the first interferometer. Using  Eq. (\ref{vis2S}), we can obtain for the visibility of the first particle
\begin{multline}
\mathcal{V}=
2\mid (r_1r_2 e^{i\phi_1})(t_1r_2 e^{-i\phi_3})+(t_1t_2 e^{-i\phi_4})(r_1t_2 e^{i\phi_2})\mid.
\label{2V}
\end{multline}
Therefore, the square of the visibility provides
\begin{eqnarray}
\nonumber
\mathcal{V}^2&=&4(r_1t_1)^2\mid r_2^2+t_2^2e^{-i\phi}\mid^2\\ 
 &=&4(r_1t_1)^2[ r_2^4+t_2^4+2(r_2t_2)^2\cos{(\phi)}].
\label{3V}
\end{eqnarray}
As can readily be seen from this result,  if the refectivity in the second interferometer becomes zero ($r_2=0$), the the gravitational interaction would have no effect on the visibility of the first interferometer.

In Fig. \ref{fig2v1} the results for visibility is plotted. In  Fig. \ref{fig2v1}(a), the visibility of the system as a function of reflectivity of first beam splitters in each interferometer is depicted, where the gravitaionally induced phase is set to $\phi=\pi$.
Fig. \ref{fig2v1}(b) illustrates the dependence of visibility on the gravitaionally induced phase and  reflectivity of the first beam splitter in the first interferometer. 

Finally we attain the predictability of the first mass from Eq. \eqref{disting2S} as
 \begin{align}
\mathcal{P}^2=(r_1^2-t_1^2)^2.
\label{1disting}
\end{align}

Considering these relations the equality in Eq.  (\ref{complementarity2s1}) is fulfilled. As a result, the gravity induces a tradeoff between  entanglement and complementarity. 

In Fig. \ref{fig2p1} the results for the predictability in terms of various involving parameters is illustrated. In   Fig. \ref{fig2p1}(a) predictability as a function of reflectivity of beam splitters in each interferometer is depicted. Here the gravitaionally induced phase is fixed to  $\phi=\pi$. In  Fig. \ref{fig2p1}(b) the dependence of predictability on the gravitaionally induced phase and  reflectivity of the first beam splitter in first interferometer is illustrated. As we expect from Eq. \ref{1disting} it only dependence on the reflectivity of the first beam splitter and for $r_1=1/\sqrt{2}$, the path information entirely vanishes.

 Taking $r_1=r_2=1/\sqrt{2}$, the entanglement becomes maximum, while visibility reduces. Practically, choosing 50:50 beam splitter observing the quantum entanglement features of the gravity becomes more feasible.  In this case, the relations for the first particle simplifies to
\begin{eqnarray}
\nonumber
 \mathcal{C}^2&=&2(r_1t_1)^2 (1- \cos(\phi)), \\ \nonumber
\mathcal{V}^2&=&2(r_1t_1)^2 (1+\cos(\phi)),\\
 \mathcal{P}^2&=&(r_1^2-t_1^2)^2.
\label{50BS2}
\end{eqnarray}
This readily provides $\mathcal{P}^{2}+\mathcal{V}^{2}+\mathcal{C}^{2}=1$. Therefore, quantum nature of the gravity provides complementarity and entanglement, enabling  a feasible experimental analysis of the quantum gravity, and providing important features of the quantum nature of the gravity.

 \section{Bell inequality and testing quantum gravity}
  \label{Bell}
 
The non-local correlation exhibited by quantum entanglement was first addressed in 1935 by Einstein, Podolsky and Rosen  in a famous paper (usually called the EPR paper) \cite{einstein1935can}, which demonstrated the apparently paradoxical context of quantum mechanics. Almost three decades later, John Stewart Bell found an experimentally manageable and quantitative platform to investigate this controversial feature of composite quantum systems and how to discriminate it from classical descriptions \cite{bell1987einstein}. Here, we discuss Bell inequality in the context of quantum gravity, which provides an experimentally testable approach for quantum nature of the gravity. 
  The most commonly used Bell inequality is the so-called Bell-CHSH inequality \cite{clauser1969proposed,horodecki1995violating}. The CHSH operator  is given by \cite{clauser1969proposed,horodecki1995violating}
$$
\hat{B}=\vec{a} \cdot \vec{\sigma} \otimes\left(\vec{b}+\vec{b^{\prime}}\right) \cdot \vec{\sigma}+\vec{a^{\prime}} \cdot \vec{\sigma} \otimes\left(\vec{b}-\vec{b^{\prime}}\right) \cdot \vec{\sigma},
$$
where $\vec{a}, \vec{a^{\prime}}, \vec{b}, \vec{b^{\prime}}$ are unit vectors.
Given the above relation, the Bell-CHSH inequality reads \cite{clauser1969proposed,horodecki1995violating}
$$
|\langle\hat{B}\rangle|=|\operatorname{tr}(\rho \hat{B})| \leq 2.
$$
 Violation of this inequality, i.e., $|\langle\hat{B}\rangle|> 2$, indicates existence of quantum correlations in a quantum state, and hence,  it provides an experimentally appealing method for the test of quanta of the gravity.
 
 As an interesting connection, there is a relation between entanglement and the maximum violation of Bell-CHSH inequality given as \cite{ghosh2001mixedness,verstraete2002entanglement}
 \begin{align}\label{bell}
\mathcal{B}=2\left(\sqrt{1+\mathcal{C}^{2}}\right).
\end{align}
 
\begin{figure}
 	\includegraphics[width=3.5 in]{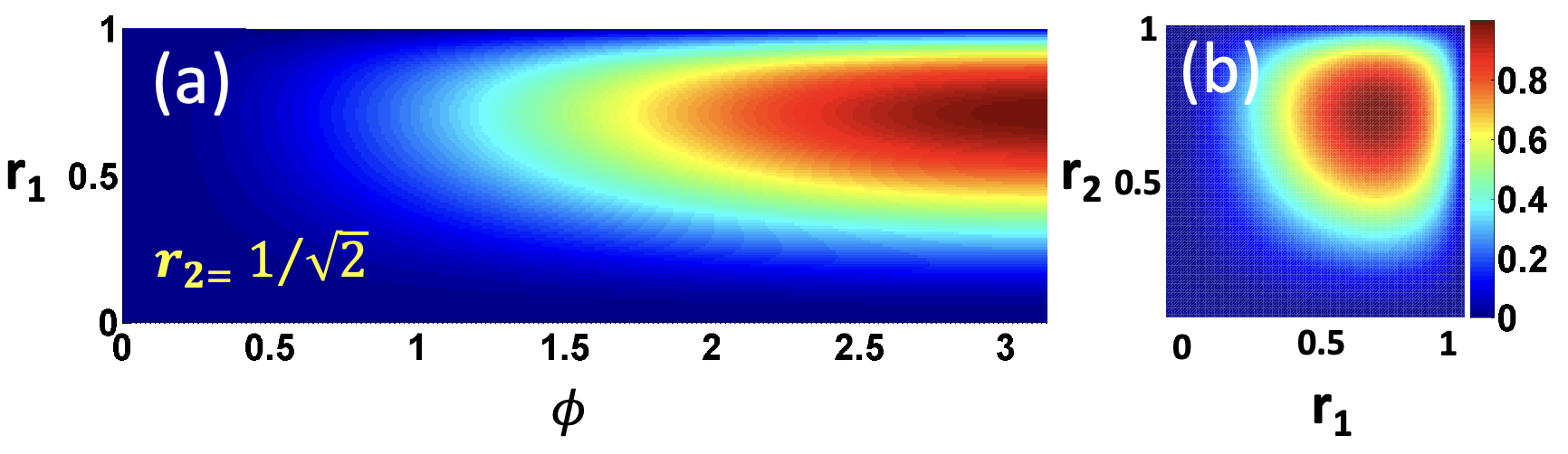}
 	\centering
 	\caption{The figure(a)  indicates the divergence of Bell parameters as function of reflectivity of the first beam splitter in the first interferometer, and the gravitationally induced phase. In the figure(a) , the divergence form Bell  parameter as a function of reflectivity of  first beam splitters of first interferometers is plotted, in which the gravitaionally induced phase is fixed to  $\phi=\pi$. }	
 		
 	\label{Fig2b1}
 \end{figure}

 This in turn shows that Bell inequality can be violated by all entangled states ($\mathcal{C}\neq 0$).
 Since the quantum mechanical non-locality  appears when this parameter exceeds two, we plot $\mathcal{I}=\mathcal{B}-2$ in Fig. \ref{Fig2b1}. Accordingly, Fig. \ref{Fig2b1}(a) indicates the violation of Bell inequality as a function of reflectivity of first beam splitter in first interferometer and the gravitationally induced phase. In Fig. \ref{Fig2b1}(b) we  demonstrate Bell parameter $\mathcal{I}$ as a function of the reflectivity of the first beam splitters in each interferometer, where we have assumed that the gravitaionally induced phase is $\phi=\pi$. Therefore, the maximum violation of the Bell-CHSH inequality occurs when both of the beam splitters are 50:50.
 
 To establish an interesting relation between complementarity and Bell parameter, using Eq. (\ref{bell}), we can attain a new relation as
 \begin{align}
 4(\mathcal{P}^{2}+\mathcal{V}^{2})+\mathcal{B}^{2}=8.
 \label{bellcomplementarity2s}
 \end{align}
 This relation demonstrates that in order to measure quantum non-locality, which is sufficient condition for the quantum nature of the gravity, we can simply measure the path-information and the visibility. This observation is  insightful for practical detection of non-classical correlations through gravity.
 
  \section{Uncertainty relations in interferometeric quantum gravity}
    \label{Uncertainty}
Now that we have considered entanglement and complementarity as the two fundamental aspects of the quantum mechanics, we now address the uncertainty principle in the context of quantum gravity, as the other fundamental concept of the quantum mechanics \cite{bagchi2016uncertainty}. The concept of quantum uncertainty was first proposed by Werner Heisenberg for the position and momentum operators \cite{heisenberg1985anschaulichen}. Later Robertson generalized it  for  any pair of non-commutative operators, such that considering a pair of non-commutative operators A and B the uncertainty relation reads \cite{robertson1929uncertainty}
 \begin{align}
\Delta{A}\Delta{B}\geq \frac{1}{2} \mid \langle [A,B] \rangle \mid,
 \label{Hur}
 \end{align}
 where  $\Delta{A}$ and $\Delta{B}$ are the standard deviations of these operators and $\langle [A,B] \rangle$ is the expectation value of the commutator of the operators.
 We could define operators  corresponding to the predictability and visibility as follows \cite{bosyk2013connection}
 \begin{eqnarray}
 \hat{ \mathcal{P}}&=&\sigma_z, \\ \nonumber
\hat{ \mathcal{V}}&=&\cos{\theta}\ \sigma_x+\sin{\theta}\ \sigma_y,  
 \label{PVop}
 \end{eqnarray}
 where
 \begin{eqnarray}
 \nonumber
  \sigma_z &=&\ket{u_1}\bra{u_1}-\ket{d_1}\bra{d_1}, \\ \nonumber
 \sigma_x &=&\ket{u_1}\bra{d_1}+\ket{d_1}\bra{u_1}, 
 \\
  \sigma_y &=& -i\ket{u_1}\bra{d_1}+i\ket{d_1}\bra{u_1}.
 \end{eqnarray}
Since the expectation values of an arbitrary operator $O$, with density matrix $\rho$, is defined as $\langle O \rangle= Tr(\rho O)$, one can write the uncertainty relation for the predictability  and the visibility operators as
 \begin{align}
  \nonumber
\Delta\mathcal{P}\Delta\mathcal{V}  & \geq 2(r_1t_1)  \\
 &\times \mid r_2^2\sin({\phi_1-\phi_3-\theta}) +t_2^2\sin({\phi_2-\phi_4-\theta})\mid.
 \label{PVuncer1}
  \end{align}
  Even though this uncertainty relation is fundamentally significant in the context of quantum mechanics,  some tighter, hence better, uncertainty relations are considered as the alternative uncertainty relations \cite{bagchi2016uncertainty,bialynicki1975uncertainty,maccone2014stronger,song2017stronger}. One useful alternative uncertainty is defined as the sum of the uncertainties of each operator. As a result, for two Hermitian unitary operators $A=\vec{a}.\vec{\sigma}$ and $B=\vec{b}.\vec{\sigma}$, where $\vec{a}$ and $\vec{b}$ are the unit vectors, simultaneously acting on the density matrix $\rho=1/2(I+\vec{r}.\vec{\sigma})$, which $\vec{r}$ is a unit vector, the uncertainty is defined as \cite{bagchi2016uncertainty}
  \begin{align}
   \nonumber
  \Delta{A}^2+\Delta{B}^2 &\geq 1+\mid \vec{a}.\vec{b} \mid^2 -2\mid \vec{a}.\vec{b} \mid \\
  &\times \sqrt{1-\Delta{A}^2} \sqrt{1-\Delta{B}^2}\mid (\vec{a}\times \vec{b}).\vec{r} \mid^2.
  \label{gPVuncer1}
  \end{align}
 Now, considering such an uncertainty in the context of the problem in hand, we attain

 \begin{align}
 \nonumber
 \Delta\mathcal{P}^2+\Delta\mathcal{V}^2 & \geq 1+4(r_1t_1)^2 
 \\ 
 &\times  \mid r_2^2\sin({\phi_1-\phi_3-\theta}) +t_2^2\sin({\phi_2-\phi_4-\theta})\mid^2
 \label{gPVuncer}
 \end{align}
 
The minimum of the sum uncertainty is one, which can be obtained when one of the operators has zero uncertainty. The second term, in the right-hand-side can never exceed one, and interestingly is quadrant of the uncertainty relation in Eq. (\ref{PVuncer1}). Therefore,  the sum uncertainty bound is more tighter than the uncertainty relation in Eq. (\ref{PVuncer1}).

In an experimental setting, one can control the quantum uncertainty effects of the gravity by controlling the gravitationally induced phases. This, in fact, provides a deep connection between the quantum mechanics and gravity from the uncertainty perspective.

 \section{experimental challenges}
 \label{experimental}
The experimental investigations of quantum gravity via induced entanglement addresses one of the most important problems of the physics. Therefore, it would be important to consider the feasibility of the induced entanglement generations via current technology.  In this section, we briefly review the important challenges that such an experiments is subjected to and discuss the the feasibility of the study. 

One important challenge for generating entanglement in this realm is that one needs  to create a large enough induced phase shift that can be measurable in the experiments.  Considering the relation for the induced phase,
$ G{m_1m_2T}/{\hbar d}$, there are three parameters that we could adjust in reaching to a measurable phase in the order of unit. The first parameter is  masses of the particles. The product of the masses must be big enough to enhance the entanglement degree.  This requires preparing a large system in a quantum superposition. This problem has intensively considered in the context of quantum mechanics, and many improvements have been made thus far \cite{leggett1980macroscopic,julsgaard2001experimental,lee2011entangling,klimov2015quantum,ockeloen2018stabilized}.   Therefore, we can consider a system  with mass of about $10^{-14}$kg a reasonable mass size, for which the possible candidates could be a massive molecule \cite{arndt1999wave,gerlich2011quantum}, micro crystals \cite{bose2017spin} or Bose-Einstein condensates \cite{helmerson2001creating,peise2015satisfying}.  The other factor in controlling the induced phase is the distance between two interferometers. In a realistic setting, if the two objects get more closer to each other than a certain limit, the other interactions such as electrostatic  interactions can dominate the gravitational interaction \cite{bose2017spin,marletto2017gravitationally}. To overcomes this challenge, it has been proposed that by using a conductor in between the two masses, hence shielding other interactions,  we can adjust the distance of the interferometer as close as about $d=1\mu m$ \cite{van2020quantum}. The last parameter that we can control is the time that each particles travels in the interferometer arms. To illustrate, if we  adjust the travel time of particles in the interferometer arms in the order of  $T=0.1$sec, the induced phase could be adjust in the order of unit and consequently the maximum entanglement of two particles through gravitational interaction would be achievable. As another practical setup, one can consider
two coupled nano-mechanical oscillators with the mass $10^{-12}$kg, and the interaction time  $T = 10^{-6}$sec that would enable the phase shifts large enough to prepare maximally entangled states with the distance $d$ in the order of a few micrometers \cite{marletto2017gravitationally}.
Therefore, in such a timescales we must protect the system from decoherence from the environment. 
To eliminate the environments effect on the system we need a high vacuum and a low temperature. The estimated pressure is in the order of $10^{-15}$pa and temperature must be about $0.1$K \cite{bose2017spin}, which is an achievable condition with the current technologies \cite{ishimaru1989ultimate}.

  \section{Conclusion}
	\label{Conclusion}
	
Quantization of the gravity	remains as one of the most important, yet extremely illusive, challenges at the heart of modern physics. It has been argued that a  direct  empirical evidence for the quantum nature of the gravity (i.e., detecting gravitons) can shed light on the charactristics of the ultimate theory of the quantum gravity. However, such a task is far beyond the current capabilities and it seems not to be achievable in terms of near future technologies. To circumvent this impasse, it was recently shown that gravitationally-induced entanglement, tailored in the interferometric frameworks, can be used to detect the quantum nature of the gravity. However, many fundamental and  empirical aspects of these schemes are yet to be discovered. Considering the fact that, beside quantum entanglement, quantum uncertainty and complementarity principles are the two other foundational aspects of quantum physics, the quantum nature of the gravity needs to manifest all of these features. In this work, we considered an interferometric setup for  testing these three nonclassical aspects of quantum mechanics in quantum gravity setting, which can shed light on the connections between gravity and quantum physics in a broader and deeper discipline. As we show in this work, all of these fundamental  features of quantum gravity can be framed and fully analyzed in an interferometric scheme. We showed the relation between gravitationally-induced entanglement and complementarity principle and investigated its features. We also developed a relation between the violation of Bell inequality as a sufficient criteria for the quantumness of the entanglement and complementarity principle.

\bibliographystyle{apsrev}
\bibliography{library}

\end{document}